\documentclass[twocolumn]{aastex63} 
\usepackage[utf8]{inputenc}
\usepackage{comment}
\usepackage{amsmath}
\usepackage{graphicx}
\usepackage{textcomp}
\usepackage{gensymb}

\newcommand{\rproc}{\textit{r}-process} 

\begin{document}
\title{On the Blackbody Spectrum of Kilonovae}

\correspondingauthor{Albert Sneppen}
\email{albert.sneppen@nbi.ku.dk}

\author[0000-0002-5460-6126]{Albert Sneppen}
\affiliation{Cosmic Dawn Center (DAWN)}
\affiliation{Niels Bohr Institute, University of Copenhagen, Lyngbyvej 2, K\o benhavn \O~2100, Denmark}


\begin{abstract}
    The early spectra of the kilonova AT2017gfo have a remarkably smooth blackbody continuum, which reveals information on the thermal properties and radioactive heating within the ejecta. However, the widespread use of a single-temperature blackbody to fit kilonova data is theoretically invalid, because 1) the significant travel-time delays for a rapidly cooling surface result in a broad distribution of temperatures and 2) the relativistic Doppler correction varies across different surface elements. Thus, the observed spectrum should be a modified blackbody with a range of temperatures over the surface. In this paper we quantify the impact of these effects and illustrate the typical wavelength-dependent spectral corrections. We apply the multi-temperature blackbody framework to the first epoch X-shooter AT2017gfo spectrum, to deconvolve the underlying physical temperature at the photosphere from the relativistic Doppler shift. We show that cooling and Doppler effects individually results in a variation of temperatures over the photosphere of up to 30\%, but in combination these effects nearly cancel and produce the single-temperature blackbody observed. Finally, we show that fitting the UV, optical or NIR separately yields blackbody temperatures consistent at the percent-level, which puts stringent limits on any proposed modification of the spectral continuum. \newline 
    
\end{abstract}

\section{Introduction}
Early kilonova spectra are often modelled with a single-temperature blackbody continuum as 
observationally validated in the early spectra taken of the well-studied kilonova AT2017gfo \citep{Drout2017,Shappee2017,McCully2017,Pian2017,Waxman2018,Watson2019}. However, even an emitted single-temperature blackbody will not be observed with a single temperature due to the mildly relativistic velocities, the significant time-delays, and the rapid cooling rate of kilonovae \citep{Sneppen2023}. This is caused by three competing and complementary effects. Firstly, at one day after the explosion, the light-travel time across the ejecta is several hours, in which time the ejected material itself cools rapidly. Therefore, the nearest front of the ejecta (which is observed at the latest time of the expansion) is cooler than more distant portions of the ejecta. Secondly, the different parts of the ejected material is Doppler boosted by different amounts depending on their varying projected velocity along the line-of-sight. Thirdly, the thermalisation depth is wavelength-dependent, which suggests photons at different wavelengths originate from different radial depths and potentially temperatures. While the wavelength-dependent opacity is well-known and discussed \citep[e.g.][]{Tanaka2020}, the spectral and statistical significance of the latitude-dependent effects have yet to be quantified.

Therefore, in Sec. \ref{sec:deri} we derive the multi-temperature blackbody and illustrate the spectral significance for typical kilonova velocities, time-scales and cooling-rates. In Sec. \ref{sec:prior}, we review and summarize the prior constraints for cooling rates and characteristic velocities for AT2017gfo. In Sec. \ref{sec:fit}, we explore the constraints attainable by modelling and fitting this multi-temperature blackbody for the 1st epoch X-shooter spectrum of AT2017gfo. Lastly, we discuss the applicability and significance of the multi-temperature frameworks for future kilonovae observations and the implications for wavelength-dependent thermalisation depths. All reported errors consistently refer to the 1$\sigma$ uncertainty


\section{Derivation of the Modified Blackbody}\label{sec:deri}
We define a coordinate-system centered at the explosion's core with the axis pointed towards the observer. Each position on the photospheric surface can then be parameterised in terms of the polar ($\theta$), azimuthal ($\phi$) and radial ($R_{\rm ph}$) coordinates from the source center. Here $\theta$ is the angle between the direction of expansion and the line-of-sight, which we will often use in the context of the projection, $\mu = \cos(\theta)$. Here, we only investigate systems with azimuthal rotational symmetry, so $R_{\rm ph}(\mu,\phi)=R_{\rm ph}(\mu)$.

Due to the increasing light travel-time to more distant parts of the ejecta, we distinguish between the passage of time as seen in the frame of the explosion centre, $t$, and the retarded time as measured by an observer, $t_{obs}$. As we measure relative to the explosion time, we will define this point in time as $t_{obs}=0$.
 
\subsection{Relativistic Doppler Corrections}
The effective temperature, $T_{\rm obs}$, (in the observers frame of reference) is boosted from the temperature in the emitted frame, $T_{\rm emitted}$, by the relativistic Doppler correction, $\delta(\mu)$. 
\begin{equation}
    T_{\rm obs}(\mu) = \delta(\mu) \ T_{\rm emitted} = \frac{1}{\gamma} \frac{1}{1-\beta \mu} T_{\rm emitted}
    \label{eq:temp}
\end{equation}

Here, $\gamma$ is the lorentz factor, $\beta$ is the velocity of the expanding photosphere, $v_{bb}$, in units of the speed of light. Naturally, the Doppler correction (and in extension $T_{\rm obs}$) varies across the different surface-elements. However, due to the varying time-delays to the rapidly cooling photosphere, $T_{\rm emitted}$ also changes across the surface, as we will detail next.

\begin{figure*}[t]
    \centering
    \includegraphics[width=\linewidth,viewport=15 15 1155 480, clip=]{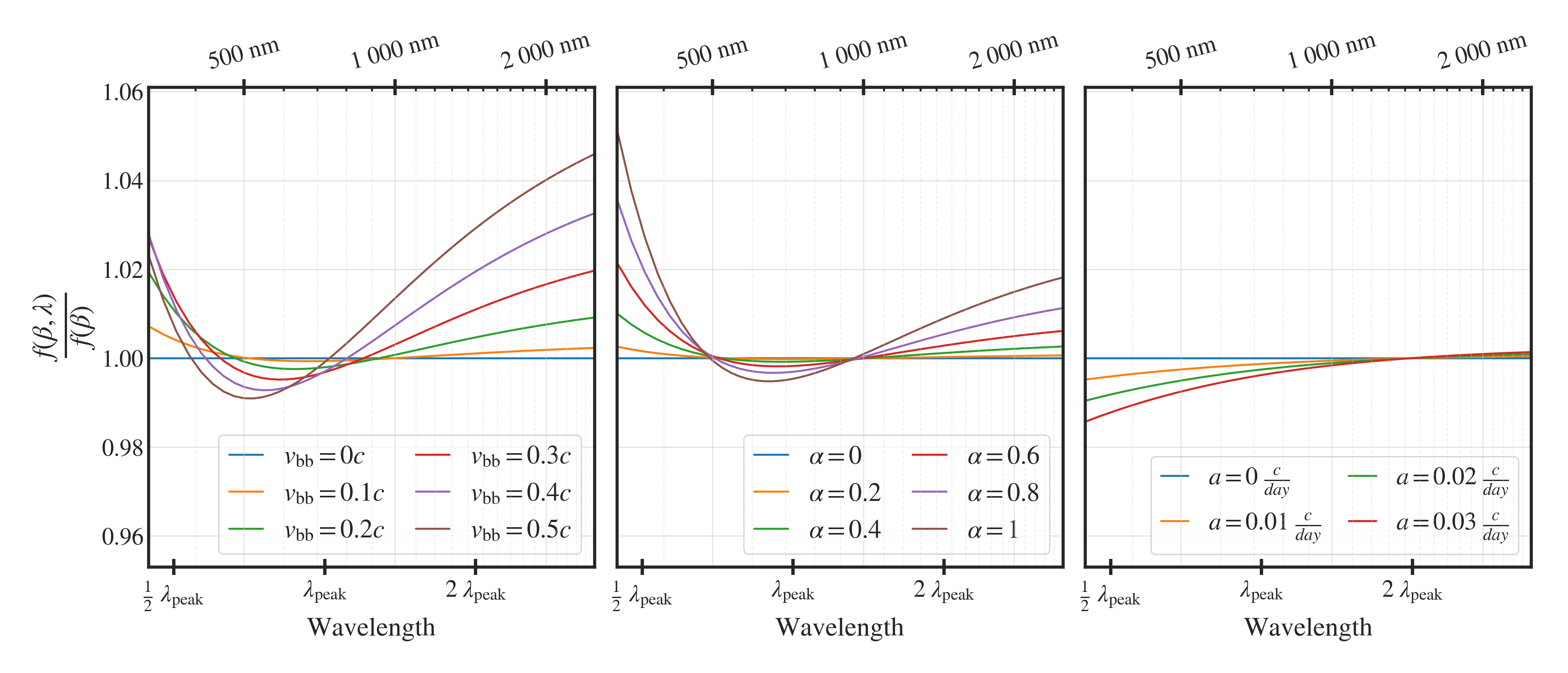}
    \caption{ The wavelength-dependence of the spectral correction, $f(\beta,\lambda)/f(\beta)$, for different velocities, $v_{bb}$ (left panel), cooling-rates, $\alpha$ (center panel), and recession-rates, $a$ (right panel). These corrections are computed by convolving blackbodies over the photospheric surface and comparing to the best-fit single-temperature blackbody. Convolving over a range of effective temperatures results in a wider blackbody with a correspondingly less prominent peak. The wavelength on the lower x-axis is shown relative to the peak-wavelength, $\lambda_{\rm peak}$, of the input blackbody, $B_{\lambda}(T_{\rm emitted}(\mu=1))$, while the top x-axis shows the corresponding wavelength assuming an emitted temperature $T_{\rm emitted}(\mu=1)=4000$\,K.
    Left: spectral correction for different blackbody velocities, while holding the velocity ($a=0$) and temperature constant ($\alpha = 0$). Center:  spectral correction given variable cooling-rates and disregarding the relativistic Doppler corrections. To compute the characteristic time-delays we use the characteristic velocity of AT2017gfo in Epoch 1; $v_{bb} = 0.28c$. Right: amplification of the spectral correction introduced by different velocity recession-rates, $a$, with a velocity at the nearest front $v_{bb}(\mu=1) = 0.28c$ and a constant temperature ($\alpha=0$).} 
    \label{fig:ratio}
\end{figure*}

\subsection{Temperature-cooling and light-travel time effects}\label{sec:temp_travel}
The temporal evolution of temperature in a kilonovae atmosphere is theoretically set by the competition between adiabatic cooling, radiative cooling and heating from decaying \rproc \ elements.  
The decays from a large statistical ensemble of isotopes is predicted to provide a heating rate well-approximated by a single powerlaw \citep{Metzger2010b}. Indeed, the observed light-curves of AT2017gfo (from 1 to 6 days post-merger) show that the bolometric luminosity is well-described by such a powerlaw-decay, $L\propto t^{-1}$ \citep{Waxman2018}, which has been interpreted as observational evidence of \rproc \ heating \citep{Wu2019}. Thus, given the blackbody nature of early emission, the standard prescription for the temperature as a function of the time since the merger is the powerlaw relation: 
\begin{equation}
    T_{\rm emitted}(\mu,t_{obs}) = T_{\rm ref} \cdot \left(\frac{t(\mu,t_{obs})}{t_{\rm ref}} \right)^{-\alpha} 
    \label{eq:cool}
\end{equation}
Here, $T_{\rm ref}$ is the reference temperature at some reference time, $t_{\rm ref}$, while $t$ encompasses the time-delay to different surface elements as set by the geometry and the physical size of the system. Eq. \ref{eq:cool} thus yields the emitted temperature, $T_{\rm emitted}$, given any powerlaw index $\alpha$ and a set of travel-times to the different surface-elements. 

The typical estimates for the powerlaw-index, $\alpha \approx 0.5$ \citep{Metzger2010b}. We emphasise, the powerlaw evolution in temperature is a theoretically-motivated prescription, which empirically breaks down in later phases but provides a decent fit across spectra in the photospheric epochs \citep{Waxman2018}. 

The arrival time for light to reach the observer from a surface-element with position ($R_{\rm ph},\mu$) and emitted at time, $t$, is: 
\begin{equation}
\begin{split}
    t_{obs} &= t - \mu R_{\rm ph} / c
\end{split}\label{eq:time-delay}
\end{equation}
Here, $t$ represents the time it takes for the photosphere to expand to the position ($R_{\rm ph},\mu$), while the second term accounts for the variations in light travel-time from the photosphere to the observer. Inserting the relation for a spherically symmetric and homologous expansion, $R_{\rm ph} = \beta c t$, and solving for $t$:
\begin{equation}
\begin{split}
    t(t_{obs},\mu) &= \frac{t_{obs}}{1 - \mu \beta}
\end{split}\label{eq:time-delay}
\end{equation}
That is light arriving from more distant ejecta (lesser $\mu$) must have been emitted earlier than light from nearer parts of the ejecta. As there will have been less time for expansion, the homologous form of the equation, $R_{\rm ph}(\mu,t_{obs}) = \beta c t_{obs} / (1 - \mu \beta )$, dictates that the photospheric radius must then be smaller near the limb of the ejecta. This so-called surface of equal arrival time for a spherical expansion of constant velocity was first derived in \cite{Rees1967}. Taking the ratio of radii we get $R_{\rm ph}(\mu)/R_{\rm ph} = (1-\beta)/(1-\beta \mu)$, where $R_{\rm ph}$ is the shorthand notation for $R_{\rm ph}(\mu=1)$, which we will use in Sec \ref{sec:deri_final} when weighing the spectral contribution from different surface-elements. 
The spherical expansion assumed in this derivation is that of the observed photosphere \citep[as constrained in][]{Sneppen2023} which is not necessarily identical to the total opacity. These two may differ as there is a larger column-density for line-of-sight from higher latitudes, $\theta$, which may shift the last-scattering surface outwards. This effect is sensitive to the exact composition, density and opacity profile, with the maximum shift being for ejecta with a relatively uniform radial structure. However, typical models suggesting a rapid radial decline in opacity and density with $\rho \propto r^{-n}$ where $n \in [3;6]$ (see for instance \cite{Collins2023}), which would suggest a $2-3$\% level difference at the half-light radius. This bias is $0.7-1$\,dex smaller than the characteristic time-delay, so the observed spherical photosphere used in this analysis is close to the underlying geometry of the ejecta. This near correspondence is observationally corroborated by the consistency in geometry over multiple epochs \citep{Sneppen2023}, whereas this limb-projection bias (if significant) in contrast should evolve in a time-dependent manner.

Naturally, one can generalize these arguments and numerically compute the time-delays for any arbitrary geometry. Indeed, in the next section we will discuss the impact of a empirically-motivated perturbation of the framework by accounting for the recession of the photosphere.

\subsection{Receding photosphere}
A notable systematic effect for this analysis is the sub-linear growth of the photospheric surface. The ejecta approaches homologous expansion, $R \propto t$, within a timescale of minutes post-merger \citep{Rosswog2014}. However, as the outer layers becomes increasingly optically thin, the photospheric surface recedes deeper into the ejecta. It is computationally straightforward to include any temporal evolution in velocity within Eqs. \ref{eq:cool} and \ref{eq:time-delay} given any parameterization of the velocity, $v_{bb}(t)$. We test a simple linear expansion in velocity around 1.4 days post merger $v_{bb}(t) = v_0 - a \cdot (t-1.4 \ \rm{days})$ where $a$ is the recession-rate of the photosphere, to model the sensitivity of this analysis to this effect. 

For AT2017gfo, the X-shooter spectra at epoch 1 and 2 (1.4$-$2.4 days post merger) both the photospheric velocity from the 1 $\mu$m P~Cygni feature \citep{Watson2019} and the velocity inferred from the blackbody fits \citep{Sneppen2023} decreases from 0.28c to 0.25c. Thus, given the X-shooter velocity constraints the expected parameter-values are around $v_0 \approx 0.28c$ and $a\approx 0.03 \frac{\rm{c}}{\rm{day}}$. For the characteristic light-travel time-scale of several hours, this effect is a noteworthy but relatively minor correction when compared to the other effects discussed (see Fig. \ref{fig:ratio}). Note, blackbody fits to earlier spectra (at 0.49 and 0.53 days post-merger) are less constraining, but are still consistent with the velocity inferred 1.4 days post-merger \citep{Shappee2017}. This may indicate the recession-rate is even smaller in the period up to the 1st epoch X-shooter spectrum.  


\subsection{The integrated expression}\label{sec:deri_final}
Ultimately, this suggests that the observed temperature is a convolution of cooling temperatures and varying Doppler corrections. To determine the corresponding spectral shape of a multi-temperature blackbody, we must integrate the blackbodies over the surface (and in extension over the range of effective temperatures). 


The specific total luminosity is $L_{\lambda} = 4 \pi D_L^2 F_\lambda$, where $D_{L}$ is the luminosity distances and \(F_\lambda\) is the wavelength specific flux. The specific flux, $F_{\lambda}$ is itself the integral of the specific intensity, $I_\lambda$, over the solid angle of the source object as seen from the observer, $d\Omega_{\rm obs}$ \citep[e.g][eq.\,1.15]{Ghisellini2013}:

\begin{equation}
    L_{\lambda} = 4 \pi D_L^2 \, F_{\lambda} = 4 \pi D_L^2 \int_{\Omega_{\rm obs}} I_\lambda \ \mu' \ d\Omega_{\rm obs}
\end{equation}
where the projection $\mu'\approx1$ as the source is distant and thus the incident rays are near perpendicular to the detector.  We insert the solid-angle as seen from the source center, $d\Omega=d\mu d\phi$, in place of the solid angle as seen from the observer using the relation $d\Omega_{\rm obs} = \frac{\mu dA}{D_L^2} = \left(\frac{\mu R_{\rm ph}(\mu)^2 d\mu d\phi}{D_L^2} \right)$, where the $dA$ is the emitting surface-element at $R_{\rm ph}(\mu)$: 
\begin{equation}
    L_{\lambda} = 4 \pi \iint I_\lambda R_{\rm ph}(\mu)^2 \mu \ d\mu d\phi \\
\end{equation}

Noting, that the expression is independent of the azimuthal angle $\phi$ due to cylindrical symmetry, the integral over this angle is trivial $\int_0^{2\pi} d\phi = 2\pi$ and for the general case $I_\lambda$ will vary across the photospheric surface, so we get: 

\begin{equation}
\begin{split}
    L_{\lambda} 
    &= 8 \pi^2 \int I_\lambda(\mu) R_{\rm ph}(\mu)^2 \mu \ d\mu \\
\end{split}\label{eq:generic}
\end{equation}

In the following, we deliberate on each of the constituent parts of this integral:

Firstly, when the radiation is in thermodynamic equilibrium, then the initial specific intensity is given by the Planck function, $I_\lambda(\mu)=B_\lambda(T_{\rm obs}(\mu))$, at the observed temperature, $T_{\rm obs}(\mu) = \delta(\mu) T_{\rm emitted}(\mu)$. It is a mathematical property of a blackbody in bulk motion, that while the relativistic transformations boost the observed luminosity, this is exactly matched by a higher effective temperature \citep[e.g][Table 3.1]{Ghisellini2013}. That is the relativistic Doppler correction of a blackbody is indistinguishable from a blackbody with a shifted effective temperature. 

Secondly, the integral over the projection $\mu$ would in the co-moving frame of the ejecta range from $0$ to 1, but the lower limit of the integral will increase to $\beta$ in an observer's frame due to the aberration of angles (as detailed in \cite{Sadun1991}). This can equivalently be interpreted as due to light travel time effects the surface of equal arrival time with the largest cross-section is at $\mu = \beta$. Anything located at $\mu<\beta$ will be hidden behind the nearer front.

Lastly, as deliberated in Sec. \ref{sec:temp_travel} for a spherical photosphere expanding with a velocity, $\beta$, the surface of equal arrival time is $R_{\rm ph}(\mu) = R_{\rm ph} ((1-\beta)/(1-\beta \mu))$, where the geometrical factor in parenthesis accounts for light arriving from the limb having been emitted relatively earlier than from the front, when the photosphere had a smaller surface area \citep{Rees1967}.

Inserting these identities in Eq. \ref{eq:generic} yields the functional form for the specific luminosity:

\begin{equation}
\begin{split}
    &L_{\lambda} = 8 \pi^2 \int I_\lambda(\mu) R_{\rm ph}(\mu)^2 \mu \ d\mu \\
    &= 8 \pi^2 R_{\rm ph} \int_{\beta}^1 B_\lambda(T_{\rm obs}(\mu)) \left(\frac{1-\beta}{1-\beta\mu}\right)^2  \mu \ d\mu \\
    &= 8 \pi^2 R_{\rm ph} \int_{\beta}^1 B_\lambda\left(\delta(\mu) T_{\rm ref} \left(\frac{t(\mu,t_{obs})}{t_{\rm ref}} \right)^{-\alpha} \right) \left(\frac{1-\beta}{1-\beta\mu}\right)^2  \mu d\mu \\
    &= 4 \pi R_{ph}^2 \pi B_\lambda(T_{\rm obs}) f(\beta, \lambda) 
\end{split}\label{eq:final}
\end{equation}
where we have first inserted the equal arrival time surface of a spherical expansion at constant velocity, then written the full projection-dependent blackbody for each surface element, and finally introduced the wavelength-dependent correction, $f(\beta, \lambda)$, to the spectral shape from relativistic and time-delay corrections. If the wavelength-dependence is small, $f(\beta,\lambda) \approx f(\beta)$, then the luminosity-weighted spectrum remains largely well-described by a single blackbody. 

While Eq. \ref{eq:final} assumes a spherical expansion with a constant speed, Eq. \ref{eq:generic} remains more generalizable as one can invoke any axially-symmetric geometry and any prescription of velocity. For example, one can include the time-delay effect of the receding velocity by including $\beta(t)$ in the expressions for the time-delay, $t(\mu,t_{obs})$ and the photospheric radii, $R(\mu,t_{obs})$. While the analytical prescription for an arbitrary geometry can quickly become protracted, it remains straightforward to numerically compute the relevant time-delay and emitting areas for any perturbations of the geometrical shape. That is given any prescribed geometry, one determines the surface-coordinates as seen by an observer by computing the radius, $R(\mu,t_{obs})$, where the photospheric surface intersects with the equal arrival time surface.  

In Fig. \ref{fig:ratio}, we illustrate the resulting change in blackbody shape caused by varying the velocities, temperatures or recession-rates of the photospheric surface given the spherically expanding photosphere observed in \cite{Sneppen2023}. The characteristic shape of these modified blackbodies have a less prominent peak and an excess of emission in the tails - ie. an increase in the spectral width compared to a standard blackbody. We note, that to detect varying Doppler corrections from the spectral shape necessitates mildly relativistic velocities. For instance, a velocity of $v_{bb} = 0.3c$ produces variations in temperature of around 30\% over the surface, but due to the outshining of the brightest blackbody-component this only results in a bias of a few percent comparing the optical with the UV or NIR. Additionally, the spectral shape is only sensitive to the spread of effective temperatures, so there is a large degeneracy between varying rates of cooling and varying the Doppler correction. For alternative ejecta geometries, one can compute the light travel-times and the emitting area as done in the right panel of Fig. \ref{fig:ratio}, where we account for the inward recession of the photosphere over time. This recession effect (and the larger velocities it implies for the more distant part of the photosphere) results in amplification of the spectral correction from the single-velocity relativistic Doppler effect. However, this is a relatively minor supplementary correction - a $\sim 0.6$\% further shift when comparing the UV/optical and optical/NIR for the characteristic recession rate found in AT2017gfo.

\section{Temperatures and Velocities of AT2017gfo}\label{sec:prior}
An initial estimate for \textit{the temporal evolution of temperature} can be found by comparing the spectral energy distribution (SED) across epochs. Fitting best-fit blackbody models to follow-up photometry suggests the effective temperatures from 0.5 to 5.5 days is consistent with a power-law decline in time $T_{\rm obs} \propto t^{-0.54 \pm 0.01}$ \citep{Drout2017,Waxman2018}. Note, this inferred rapid cooling observed in effective temperature is due to both the intrinsic cooling of rest-frame temperature and the bias of decreasing expansion velocities, which lead to progressively less blueshifted ejecta. The photospheric approach is further biased as it precludes any contribution of emission and absorbtion lines which become increasingly prominent at later times. For these reasons, the X-shooter spectra temperatures presented in \cite{Smartt2017} and \cite{Pian2017} provides the as-of-yet most comprehensive estimate of the evolution of blackbody temperatures for AT2017gfo. Emerging features can be explicitly parameterised and fit, while the rest-frame temperature can be inferred using the observed temperature and velocity-constraints from the 1$\mu$m P Cygni feature \cite[see][]{Watson2019}. Fitting across epochs from 1.4 to 5.4 days after the merger, suggest a less-steep powerlaw with the rest-frame temperature decline being roughly consistent with $T_{\rm emitted} \propto t^{-0.46 \pm 0.05}$. The slower rate of cooling for the the effective temperature is because the contribution of the decreasing Doppler corrections is deconvolved, while the overall constraining power is significantly reduced as these high-quality X-shooter spectra are rarer and taken over a smaller dynamical range in time. 
Given the characteristic size and time-delays for the epoch 1 spectra, these cooling-rates suggest temperatures variations of order 500 K over the observed surface. 

\begin{figure}[t]
    \centering
    \includegraphics[width=\linewidth,viewport=10 10 605 450, clip=]{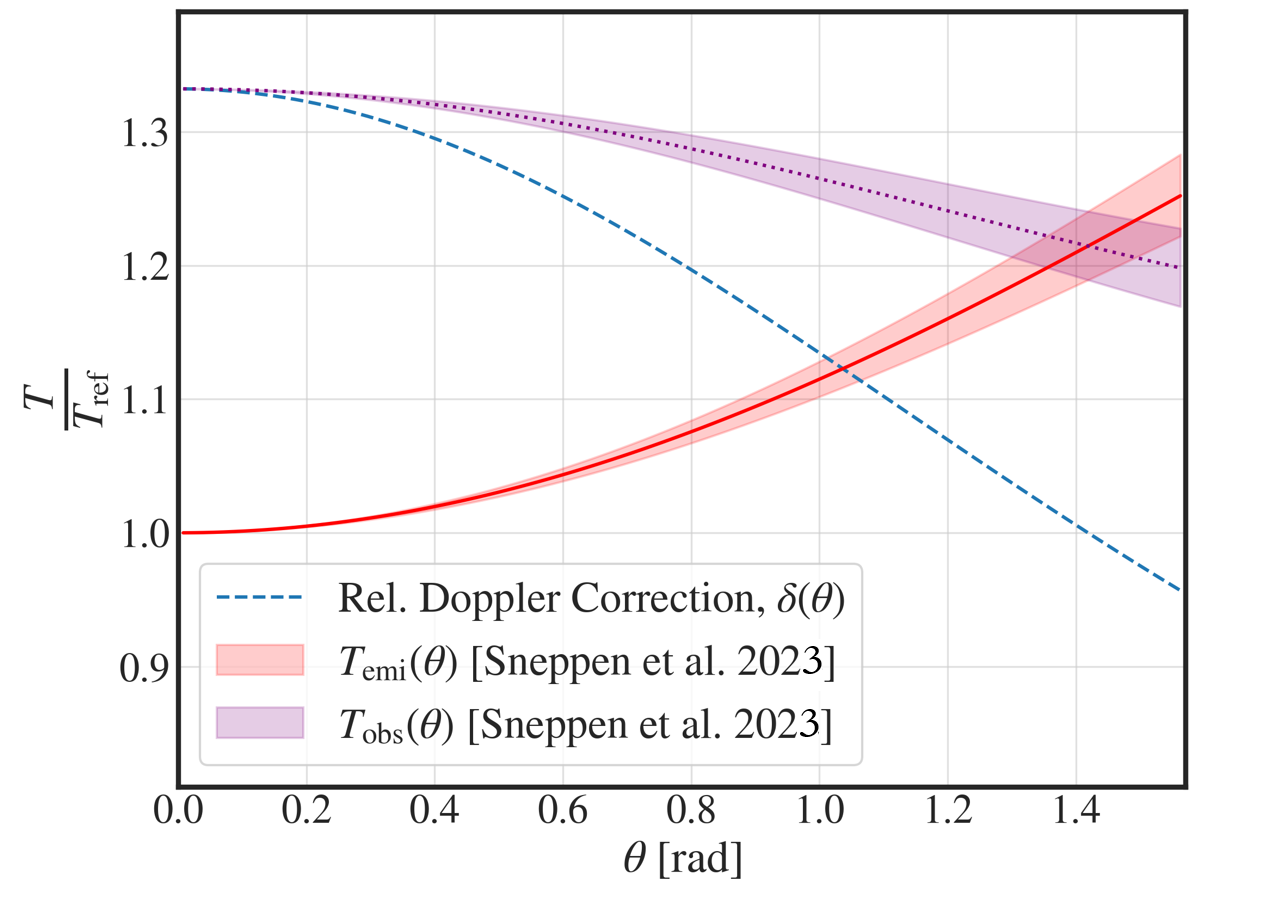}
    \caption{Temperature at the photospheric surface as a function of angle, $\theta$, between the line-of-sight and the direction of expansion for any surface element. Blue indicates the relativistic Doppler correction with $\beta=0.28$. Red illustrates the impact of time-delay showing the temperature in the emitted frame relative to the front. This is computed assuming the temporal time-scale of epoch 1 (1.43 days) and the cooling rate derived from fitting across epochs 1-5 from X-shooter spectra \citep{Sneppen2023}. Purple dotted line indicates the combined effect, which is relatively invariant across surface-elements.}
    \label{fig:rel_temp}
\end{figure}

There are several estimators of \textit{the typical velocities} in AT2017gfo. Firstly, given a known distance the overall luminosity is set by the cross-sectional emitting area of the blackbody. This area can (using the time since the GW signal) be converted into the typical cross-sectional velocity, which as first computed in \cite{Drout2017} for epoch 1 is around $0.3c$. The most recent and tight analysis which uses the improved peculiar velocity constraints for distance in \cite{Mukherjee2021}, found the cross-sectional velocity to be $v_{\perp} = 0.285 \pm 0.012$ \citep{Sneppen2023}. In addition, the identification of a Sr\(^+\) P~Cygni lines in \cite{Watson2019} allows a spectral fit of the velocity of the line, which yields a photospheric velocity of $v_{ph} = 0.278 \pm 0.001 \ c$. The velocity from the Sr\(^+\) lines are also corroborated with the velocity inferred from the Y\(^+\) P~Cygni at 0.76$\mu$m, which emerges 3-4 days post-merger \citep{Sneppen2023_Yttrium}. We note, the small statistical uncertainty of the line-velocity estimate does not include the large systematic uncertainties associated with line-blending and reverberation effects. Nevertheless, across these many prior analysis the characteristic velocity suggests the Doppler boost will incoherently shift the effective temperature of different surface elements by up to 30$\%$, which would produce a significantly wider SED than a pure blackbody spectrum. 

Curiously, the progressively increasing Doppler Correction from the orthogonal to the line-of-sight is largely off-set by the rapid cooling (see Fig. \ref{fig:rel_temp}). That is for the characteristic velocities \citep{Sneppen2023} and cooling-rates \citep{Drout2017} for AT2017gfo the variations with wavelength, $f(\beta,\lambda)$, are below 1\% over the entire spectral range. This is merely a co-incidence of the characteristic velocities, the time-scales and the rate of cooling, which combine to yield a remarkable coherent observed effective temperature across the observable surface.

\begin{figure}[t]
    \centering
    \includegraphics[width=\linewidth,viewport=25 25 485 505, clip=]{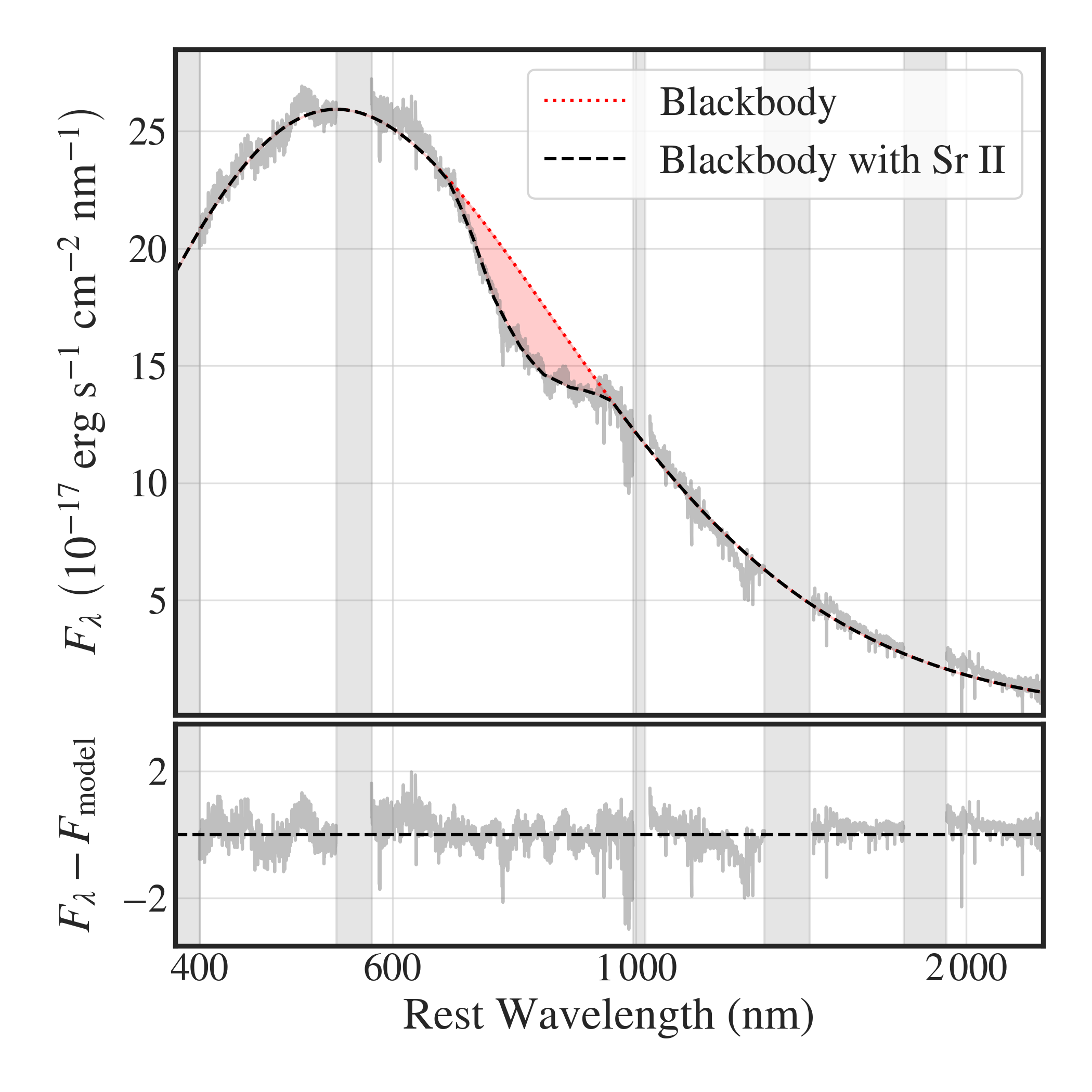}
    \caption{ Spectra of AT2017gfo 1.43 days after the merger. Spectra are from the VLT/X-shooter spectrograph (grey), with best fit blackbody and Sr\(^+\) P~Cygni. Shaded bars indicate tellurics, ill-constrained or noisy regions at the edge of the UVB, VIS and NIR arms of the spectrograph. Wavelength from 330--400\,nm are not included as spectral modelling is sensitive to the ill-constrained line-blanketing from Y\(^+\), Zr\(^+\) and other \rproc \ elements  \citep{Gillanders2022,Vieira2023}.}  
    \label{fig:1stepoch}
\end{figure}

\section{Fitting the Spectral Continuum of AT2017gfo}\label{sec:fit}
A single-temperature blackbody only contains the information of two parameters; the blackbody temperature and normalisation (which are examined in Sec. \ref{sec:intrinsic} and \ref{sec:normalisation} respectively). 
However, the broader continuum shape of a multi-temperature blackbody (such as parameterised in Eq. \ref{eq:final}) contains additional constraints on the physical effects broadening the spectral shape. Thus, fitting the blackbody width allows constraints on the physical processes perturbing the spectral shape - including the cooling-rate and the relativistic Doppler corrections (as shown in Sec. \ref{sec:mod_width}). The parameters are fittable for an observed spectrum containing well-constrained spectral lines. The closest data-set within current observations to these requirements is the first epoch spectrum of AT2017gfo (see Fig. \ref{fig:1stepoch}) taken with the X-shooter spectrograph mounted on the Very Large Telescope at the European Southern Observatory \citep{Pian2017,Smartt2017}. In this analysis, we follow the data reduction procedures outlined in \cite{Watson2019}.

This spectrum has a distinct blackbody peak separate in wavelength from the P~Cygni spectral component which in \cite{Watson2019} is associated with three Sr\(^+\) lines at 1003.9, 1033.0 and 1091.8 nm. We note another potential interpretation is that this feature originates from the Helium 1083.3 nm line, which as shown in \cite{Perego2022} out of local thermodynamic equilibrium with He masses significantly greater than the fiducial values of their models could reproduce the observed feature. As discussed in \cite{Tarumi2023}, strong 
non-LTE conditions may be present in KNe and the fraction of He produced in neutron star mergers varies greatly between numerical models, so this identification cannot currently be ruled out. 
For this analysis one can include nuisance parameters to model the 1 $\mu$m P~Cygni and the tentative emission lines at 1.5 $\mu$m and 2 $\mu$m or entirely exclude the wavelengths affected by all these lines; the constraints are robust and not sensitive to parameterizing the features or excluding these wavelengths. The optical transition-lines of Y\(^+\) producing the 0.76$\mu$m feature discussed in \cite{Sneppen2023_Yttrium} are rather weak and therefore do not affect the continuum constraints. However, the observed line-blanketing at the edge of the spectrographs sensitivity at around 350\,nm (likely due to Y\(^+\) and Zr\(^+\), see \cite{Gillanders2022,Vieira2023,Sneppen2023_Yttrium}) suggests the UV spectral shape is highly degenerate with the modelling of the UV line-opacity. Thus, 
we focus the analysis on wavelengths $\lambda\geq400\,$nm, where the spectrum is constrained. 

The fitting procedure employs a $\chi^2$-minimization to estimate the goodness-of-fit. This notable assumes the errors are reasonable approximated as Gaussian, which is following the established convention when analysing the reduced AT2017gfo spectra \citep[e.g.][]{Watson2019}. To ensure robust exploration of the parameter-landscape, we additionally sample the posterior probability distributions of parameters with Markov chain Monto Carlo \citep[using emcee;][]{Foreman-Mackey2013}, where we assume flat priors on all parameters. The peaks of these sampled posterior distributions are well centred implying that the best-fit values are well constrained. Using these fitting frameworks reveals several interesting constraints on the continuum properties detailed in the following sections. 

\subsection{The intrinsic temperature}\label{sec:intrinsic}

\begin{figure}[t]
    \centering
    \includegraphics[width=\linewidth,viewport=25 25 560 640, clip=]{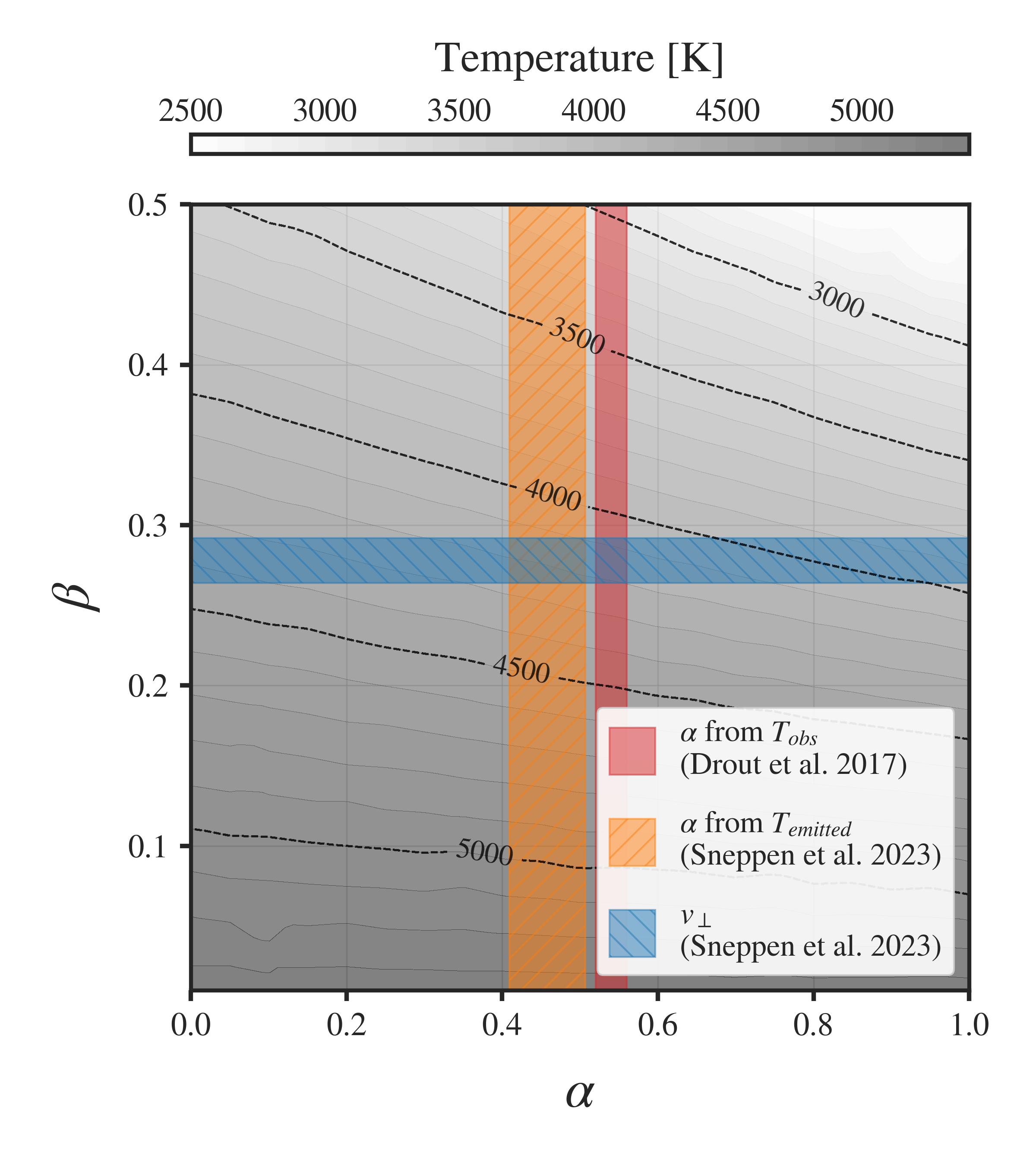}
    \caption{ The inferred temperature at the front of the expanding photosphere given varying cooling-rates, $\alpha$, and velocities, $\beta = v_{bb}/c$. Larger velocities result in a larger Doppler-corrections to the observed temperature, which naturally implies that the ejected material was cooler in the emitted reference frame. Larger cooling-rates imply the more distant portions of the photosphere are relatively hotter, so a smaller temperature at the front is required to fit the data. Constraints on $v_{bb}$ (blue, \cite{Sneppen2023}), $\alpha$ (orange and red, \cite{Drout2017}) are elaborated in Sec. \ref{sec:prior}. } 
    \label{fig:temperature_land}
\end{figure}

The observed temperature, i.e. the combination of {the intrinsic temperature}, $T_{emitted}$, and the relativistic Doppler corrections, can be deconvolved for any velocity, $v_{bb}$ (see Fig. \ref{fig:temperature_land}). As quantified in Sec. \ref{sec:prior} for the characteristic cooling-rates and velocities of AT2017gfo the variations in the effective temperature across the surface remains small. In contrast, the temperature in the emitted frame varies significantly across the surface due to light-travel time effects. Naturally, larger characteristic velocities implies lower intrinsic temperatures, but the decline in temperature with velocity is slower than the often used $T_{\rm emitted} = T_{\rm obs}/\delta(\mu=1)$, which does not account for the full luminosity-average over the visible surface. We note, given the strong constraints on the cooling-rate and velocities, one can determine the underlying physical temperature, which is essential for understanding the ejecta, the heating process and the local thermodynamic equilibrium. Combining the velocity and cooling-rate derived from X-shooter spectra \citep{Sneppen2023}, we determine the emitted temperature for the ejecta nearest the observer, $T_{\rm emitted}(\mu=1,t_{obs} = 1.4 \ {\rm days}) = 4150 \pm 60$\,K and for the ejecta furthest from the observer $T_{\rm emitted}(\mu=\beta,t_{obs} = 1.4 \ {\rm days}) = 4900 \pm 70$\,K. These temperatures are consistent with the existence of features from Sr\(^+\) \citep[eg.][]{Watson2019}, Y\(^+\) and Zr\(^+\) under LTE conditions \citep{Gillanders2021,Vieira2023}.

\begin{figure*}[t]
    \centering

    \includegraphics[width=0.49\linewidth]{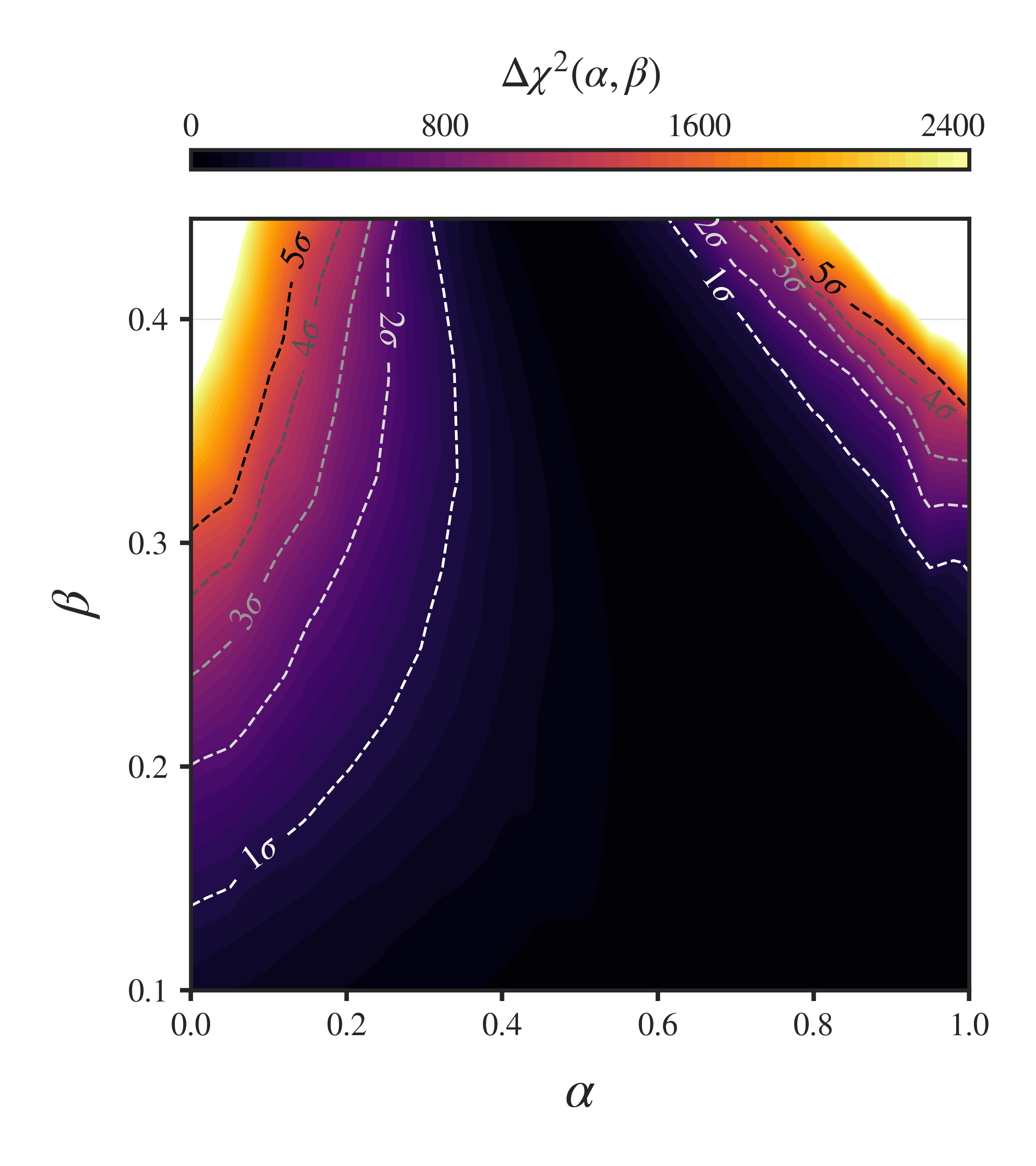}
    \includegraphics[width=0.485\linewidth]{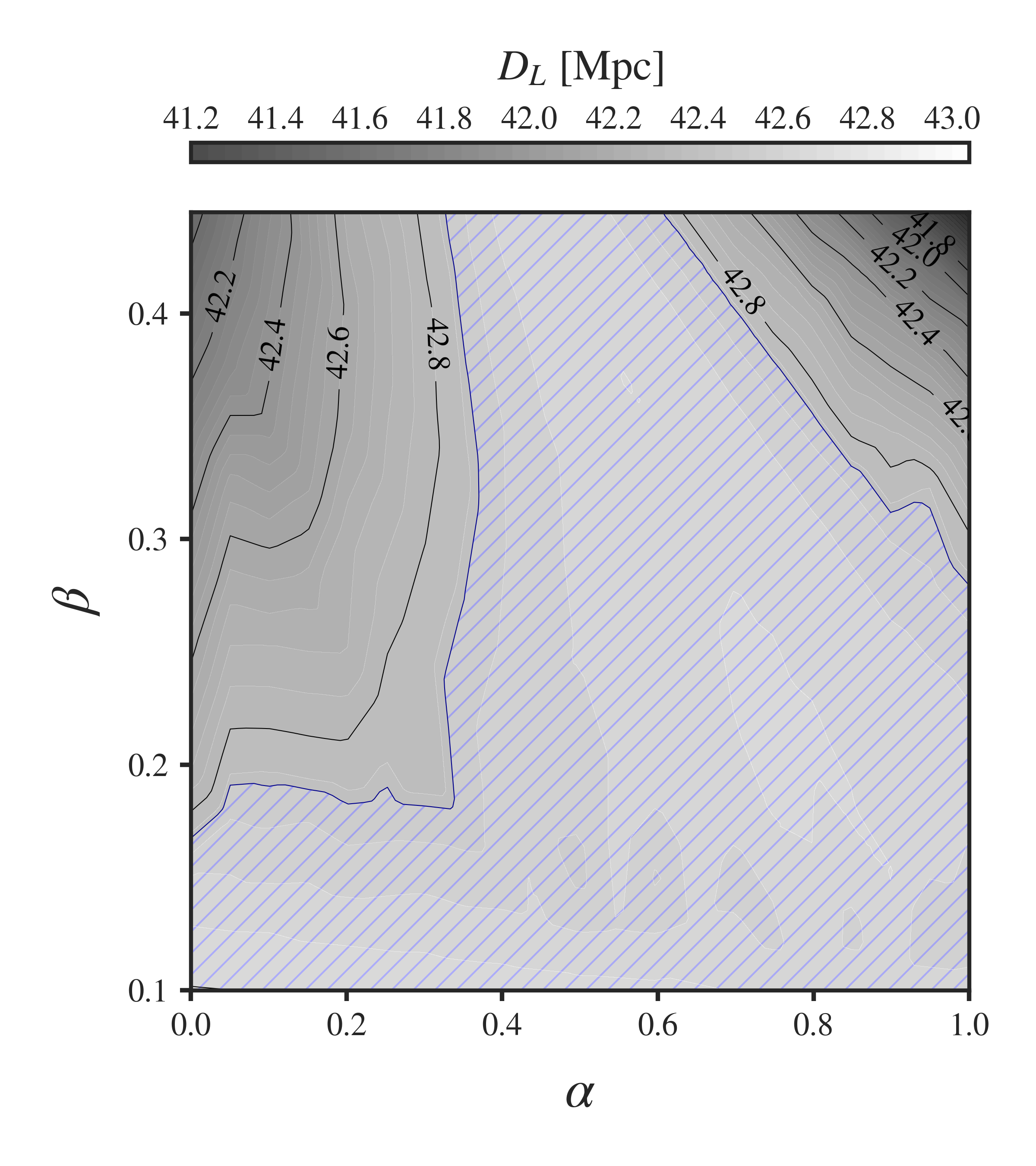} 
    \caption{Best fit $\Delta \chi^2 (\alpha,\beta)$ and luminosity distance given varying rates of cooling and velocities using the Nelder-Mead method of $\chi^2$ minimization. For each combination of parameters $\alpha$ and $\beta = v_{bb}/c$ we determine the best fit blackbody temperature, distance and $\chi^2$. 
    The dashed-lines in the left panel shows the significance level in standard deviations at which parts of the parameter-landscape are disfavored by when compared to the best-fit model. The significance level is computed from the p-value of the $\chi^2$-distribution with the relevant number of degrees of freedom when assuming $\chi_{\rm reduced}^2 =1$ for the best-fit model. 
    The blue hatched region in the right panel indicates all parts of the parameter-landscape yielding distances within $2\sigma$ of the reported distance in \cite{Sneppen2023}, $D_L = 44.5 \pm 0.8$ Mpc. Naturally, if the normalisation was kept fixed across the parameter-landscape the variation in $\chi^2$ would be even more significant.
    There are small-scale wiggles in the $\Delta \chi^2(\alpha,\beta)$ and $D_L$ landscape, because of the numerical uncertainties in finding the best-fit global minimum (and not a nearby local minimum), while there are larger monotonic trends due to the spectral corrections from $f(\beta,\lambda)$ towards low/high cooling-rates and/or higher velocities.}  
    \label{fig:landscape}
\end{figure*}

\subsection{The width of the modified blackbody: \newline Cooling and relativistic Doppler corrections}\label{sec:mod_width}
The modified blackbody width can be fit for, which allows constraints on the range of effective temperatures observed in a single epoch. For instance, the observed near-perfect blackbody continuum of AT2017gfo immediately suggests that parts of the parameter-landscape are unlikely. A rapidly expanding photosphere, $v_{bb} \approx 0.4-0.5c$, a constant temperature, or a more rapid cooling than observed for AT2017gfo, would all produce a wider blackbody and thus be disfavored by the best-fit $\chi^2$. In contrast, there are two scenarios which result in small spectral continuum deviations with $f(\beta,\lambda) \approx f(\beta)$. That is 1) the characteristic velocities are small (resulting in shorter timescales for cooling and smaller Doppler corrections), or 2) the velocities and cooling-rates offset each other. In the case of AT2017gfo, the characteristic velocity and cooling-rates inferred from spectral-fits across different epochs (see Sec. \ref{sec:prior}) are exactly in this later regime. However, given the diversity of kilonovae properties in numerical simulations, specifically in terms of cooling-rates and characteristic velocities \citep{Bauswein2013}, future objects may reside in parts of the parameter landscape, where these latitudinal effects do not cancel exactly.
 
We illustrate the constraints on the spectral shape in the left panel of Fig. \ref{fig:landscape}, showing the difference in goodness-of-fit, $\Delta \chi^2(\alpha,\beta) = \chi^2(\alpha,\beta) - {\rm min[ \chi^2(\alpha,\beta)] }$, ie. the difference in $\chi^2$ between any combination of fixed parameters and the best-fit over the entire parameter landscape. Due to the high signal-to-noise and more than 40,000 unique wavelengths in the spectrum the increase in $\chi^2$ from modifying the blackbody is noteworthy. For smaller velocities the blackbody shape provides increasingly less stringent constraints. In contrast, with increasing velocity both time-delays and Doppler variations grow, which result progressively larger deviations from a single-temperature blackbody and a smaller region of parameter landscape where the two angular effects cancel.

While the statistical constraints in fitting the multi-temperature blackbody are significant, we urge caution when interpreting only the statistical uncertainty. Modelling is limited by as-of-yet undiagnosed spectral features as such systematic effects may also create deviations from a simple blackbody and are ill-constrained without a prior model of what spectral features are present. Additionally, as mentioned previously, not only angular but also radial variations in temperature can produce a broadening of the blackbody. If future observations suggest a multi-temperature blackbody it will be ill-constrained from a single spectrum whether this is caused by radial or angular temperature-variation. However, with spectra across multiple epochs constraining the cooling-rate, the latitudinal effects can be isolated as done in this analysis.

The single-temperature blackbody observed in AT2017gfo suggests limits on both radial and angular variations in temperature are attainable (which is detailed further in Sec. \ref{sec:4.4} and \ref{sec:discussion}).

\begin{figure*}[t]
    \centering
    \includegraphics[width=\linewidth,viewport=25 67 850 415, clip=]{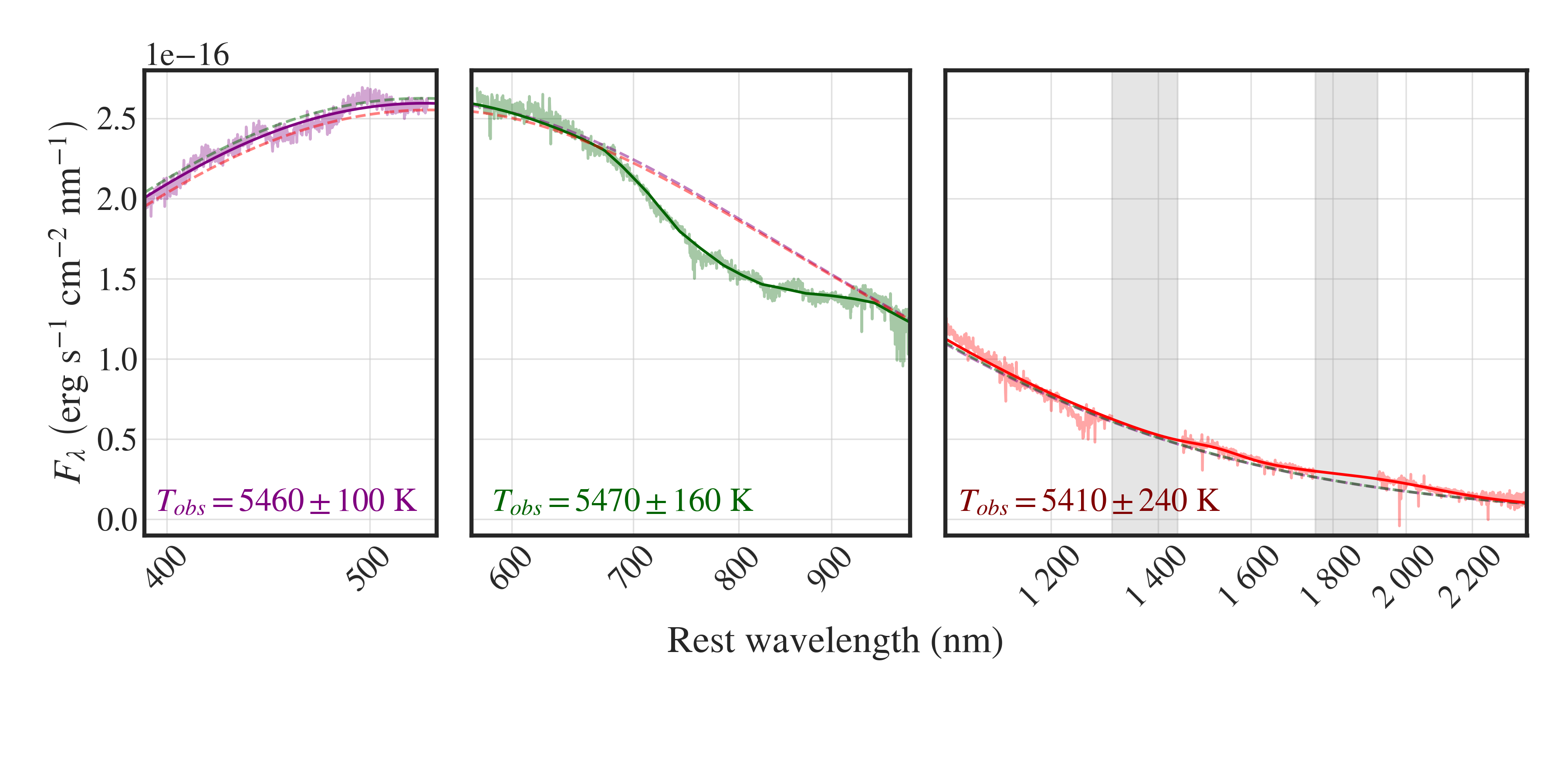}
    \caption{ X-shooter spectra of AT2017gfo at 1.43 days post-merger displaying the UV (purple), VIS (blue) and NIR-arm (red) separately. For each arm of the spectrograph, the best-fit temperature is fit independently assuming the same blackbody normalisation (ie. the same angular size). These best-fits are drawn as solid lines across the spectral-arm where they are fit, and shown as dashed lines over the wavelengths, where the fit is being extrapolated. 
    The temperature constraints fitted from each arm are consistent at the percent-level and the fits are in near-agreement with the observed spectrum even at the extrapolated wavelengths. The blackbody normalisation assumed for all epochs follows from the Planck cosmological distance $D_L=44.2\pm2.3$ Mpc to NGC\,4993 \citep{Planck2018}, and assuming the emitting area predicted from the 1$\mu$m P~Cygni \citep{Watson2019}. We also fit the proposed 1.5 and 2\,$\mu$m emission lines \citep{Gillanders2023} when constraining the NIR-arm, but ultimately these features are not prominent enough to significantly bias the best-fit temperature.}  
    \label{fig:seperate}
\end{figure*}

\subsection{The normalisation}\label{sec:normalisation}
{The blackbody normalisation} is determined by the emitting area and the distance to the source. 
Therefore given any prior constraints on the emitting area \cite[e.g.][]{Sneppen2023}, one can translate any normalisation to an inferred distance. As shown in the right panel of Fig. \ref{fig:landscape}, this implies that the best-fit distance varies across the parameter-landscape. Note, one could also define the emitting area directly from the expansion velocity of the blackbody and the time since merger, which would mainly emphasise the generic relationship between emitting area and inferred distance not the subtler dependence on the normalisation parameter. Using a strong prior on the cosmological distance [such as provided by surface brightness fluctuations \citep{Cantiello2018}, the fundamental plane \citep{Hjorth2017} or the expanding photosphere method \citep{Sneppen2023_H0}], one could further constrain the cooling-rate $\alpha$, the velocity $\beta$ or $T_{\rm emitted}$. That is that underlying parameters of the blackbody must not only satisfy a goodness-of-fit but also fall within a limited cosmological range in distance. For instance, the parts of the parameter-landscape which yield luminosity distances within 2$\sigma$ agreement of $D_L=44.5 \pm 0.8$ reported in \citep{Sneppen2023_H0} are indicated by the blue hatched region. However, given the sizeable statistical and systematic uncertainties in measuring cosmological distances and the relatively small variations in the inferred distance found across the parameter-landscape, invoking a prior on $D_L$ is unlikely to provide tight constraints on other parameters.



\subsection{The single-temperature blackbody of AT2017gfo}\label{sec:4.4}
While the proposed model (e.g. Eq. \ref{eq:final}) modifies the blackbody in a specific manner accounting for time-delays, rapid cooling and relativistic Doppler effects, we will now discuss whether other and potentially more general modifications of the blackbody are allowed by observations.  

An immediate robustness-test of the single-temperature blackbody nature of AT2017gfo (suggested in Sec. \ref{sec:mod_width}) would be to split the spectrum in different wavelength-bins and test the consistency in properties inferred for these separate parts of the spectrum. The blackbody temperature and normalisation (ie. the transients angular size) in conjunction determine $F_{\lambda}$, while the temperature is observationally constrained by the location of the blackbody-peak. Thus, fitting the blackbody normalisation and temperature concurrently necessitates wavelength-intervals near this peak. Indeed fitting the UV-arm and VIS/NIR-arms separately (ie. the wavelengths below and above the peak) yields temperatures consistent within 2\% suggesting a consistent single-temperature at least from the UV into the optical. 

This consistency in temperature can be probed across a wider wavelength-range by breaking the degeneracy with the blackbody normalisation. That is the best-fit temperature can be determined from the flux at any wavelength-band (even far from the peak wavelength) by imposing a prior on the blackbody normalisation (ie. the angular size). The angular size can be physically constrained from the measurements of the photospheric radius (such as from the Sr\(^+\)-line \citep[see][]{Watson2019,Sneppen2023}) and the distance to the transient and its host-galaxy, NGC\,4993. In the following, we will assume the Planck cosmological distance $D_L=44.2\pm2.3$ Mpc to NGC\,4993, which follows from $H_0 = 67.4 \pm 0.5 {\rm \ km \ s^{-1} \ Mpc^{-1}}$ \citep{Planck2018} and the cosmic recession velocity of $z_{cosmic} = 0.00986 \pm 0.00049$ derived from the peculiar velocity estimates in \cite{Mukherjee2021}\footnote{We note, one can equivalently assume the smaller SH0ES distance of $D_L=40.7\pm2.2$ to NGC\,4993 using $H_0 = 73.04 \pm 1.04 {\rm \ km \ s^{-1} \ Mpc^{-1}}$ \citep{Riess2021}, which implies a less-luminous transient and therefore a lower best-fit temperature for each arm. However, the best-fit temperature would still be consistent across the spectral-arms.}. Assuming this angular size is applicable across the wavelengths probed, we constrain the best-fit observed temperature for each arm independently as shown in Fig. \ref{fig:seperate}.
The best-fit model of the NIR spectral arm includes two Gaussian lines to fit the emission features at 1.5 and 2\,$\mu$m, which are observed to become prominent in later epochs \citep[e.g.][]{Gillanders2023}. However, accounting for these features ultimately proves unimportant for this analysis as fitting the NIR-arm solely as a blackbody without additional features yields a similar best-fit temperature, $T_{obs} = 5540\pm280$\,K.
The best-fit temperatures from each spectral-arm are consistent at the percent level, which results in the remarkable coherence with the observed spectrum even beyond the wavelengths of the fit. For instance, the extrapolated best-fit from the UV-arm still traces the observed spectrum at 2.2$\mu$m, while the extrapolated NIR-fit predicts the location of the blackbody peak with percent-precision. This shows that whatever perturbations of the continuum model one may propose, the derived model must still be able to explain the remarkable coherence of the data - leaving little territory for modifications of the blackbody.

We note, if one generalises these argument so that the blackbody normalisation (ie. the angular size) evolves significantly across the wavelengths/opacities probed, this imposes very strict limits on the allowed evolution in temperature. That is to match the single-temperature blackbody observed, $B_{\lambda}(T)$, while varying the normalisation with wavelength, $N(\lambda)$, would require a finely-tuned temperature, $T(\lambda)$. This fine-tuning would have to encompass a range of wavelengths (and thus an evolving sensitivity to temperature), from the linear scaling of flux with temperature in the Rayleigh-Jean's Tails (ie. $B_{\lambda \gg h c /(k_bT)}  \propto T$), over the ratio of Wien peaks fluxes (where $B_{\lambda_{\rm peak}} \propto T^{5}$), to the exponentially growing discrepancy between Wien tails. We emphasise across the UV and optical - where both normalisation and temperature can be fit concurrently - there is no observational evidence suggesting an evolution of either with wavelength. 






\section{Discussion}\label{sec:discussion}
We have derived and quantified the corrections on the spectral continuum of kilonovae due to relativistic Doppler and time-delay effects. These corrections are essential for determining the intrinsic physical temperature in the ejecta, as shown in this analysis for the epoch 1 X-shooter spectrum. More broadly, the multi-temperature blackbody framework provides a new avenue for constraining properties for kilonovae with high SNR spectra spanning ultraviolet, optical and near-infrared wavelengths. These high signal-to-noise spectra covering a broad dynamic range of wavelengths are required to constrain the percent-level continuum variations for the mildly relativistic velocities. 

Interestingly, while the standard framework of fitting single-temperature blackbodies is at least theoretically invalid for relativistic and/or rapidly cooling ejecta, it still provides a convenient functional approximation and proves remarkably coherent for the characteristic velocities and cooling-rates of AT2017gfo. We note the multi-temperature perspective is interesting in the context of future kilonova detections for probing their potentially diverse properties, where the blackbody width may provide independent and corroborating constraints. Follow-up spectroscopy of future kilonovae will peer into even earlier phases of the expansion, potentially exploring higher velocities, $v_{bb} \approx 0.4c-0.5c$, as suggested in some hydrodynamical merger simulations \citep{Bauswein2013}. Given the strongly non-linear relation between the expansion velocity and the magnitude of SED variations seen in Fig. \ref{fig:ratio}, such observations may deviate from a simple blackbody model. 
 
Lastly, we note that the observational practice of fitting a Planck function to the kilonova spectra is remarkably empirically accurate especially when compared to current radiative transfer simulations \citep[eg.][]{Shingles2023}. For instance, the wavelength-dependent opacity of the atmosphere and the temperature-gradients within the ejecta may naively imply the need for additional spectral corrections as different radial temperatures introduce an additional broadening of the blackbody. As such, any prescription for the opacity and temperature-gradient can easily be modelled and included within the framework of this analysis (alongside the other parameters investigated). However, modelling is complicated as the opacity is dominated by lines from \rproc \ elements, which currently have limited atomic data and incomplete line-lists \citep{Gaigalas2019,Tanaka2020,Gillanders2021}. 


The significant wavelength-dependent opacity of kilonova atmospheres suggests different wavelengths probe different radii \citep{Kasen2013}, which, if combined with a radial temperature gradient, would suggest different wavelengths are probing different characteristic temperatures (ie. the observed spectrum is a convolution of radial blackbodies). A convolution of either radial and angular temperature variations will produce a broader blackbody, while the combination of radial and angular variations will be larger than either effect individually. Therefore, the single-temperature nature of the observed blackbody necessitates 1) the uniform effective temperature over the observed latitudes of the photosphere, but it also requires 2) a rapid transition to optical thickness as the temperature at the thermalisation depth must be relatively consistent across wavelengths \citep{Pe'er2011}. This analysis illustrates that the angular variations in $T_{\rm obs}$ are small for AT2017gfo, but to reproduce the observed single-temperature blackbody also requires a small variation of $T_{\rm obs}$ over the radial range of thermalisation depth. For instance, fitting the UV, optical and NIR separately yield blackbody temperatures consistent at the few percent level, indicating a consistent single temperature at least at the optical depths probed from the UV into the near-infrared (see Sec. \ref{sec:4.4}). Any further quantified limits require follow-up research to both constrain the wavelength-dependent thermalisation depth and the nature of the gradient in temperature. Such discoveries may provide additional insights into the origin of the blackbody spectrum and the environment of kilonova atmospheres. \newline

The author would like to thank Stuart Sim and Ehud Nakar for independently recreating the central derivations and Darach Watson for feedback and comments on the manuscript. Additional gratitude is expressed towards Oliver Just and Andreas Flörs for useful discussions. I would also like to express my appreciation for the anonymous referee for the substantial, detailed and diverse comments. The Cosmic Dawn Center (DAWN) is funded by the Danish National Research Foundation under grant No. 140. 





\section*{Data Availability}
Work in this paper was based on observations made with European Space Observatory (ESO) telescopes at the Paranal Observatory under programmes 099.D-0382 (principal investigator E. Pian), 099.D-0622 (principal investigator P. D’Avanzo), 099.D-0376 (principal investigator S. J. Smartt) and 099.D-0191 (principal investigator A. Grado). The data are available at http://archive.eso.org.

\bibliographystyle{aasjournal}
\bibliography{refs.bib} 

\end{document}